\documentclass[a4paper,fleqn,usenatbib]{mnras}

\usepackage{txfonts}

\usepackage[T1]{fontenc}
\usepackage{ae,aecompl}

\usepackage{graphicx}	

\newcommand{\rosat}{{\sl ROSAT\/}}

\newcommand{\swift}{{\sl Swift\/}}

\newcommand{\nh}{{N$_{\rm H}$}}
\newcommand{\epcs}{ergs\,cm$^{-2}$s$^{-1}$}
\newcommand{\eps}{ergs\,s$^{-1}$}
\newcommand{\cps}{ct\,s$^{-1}$}
\newcommand{\kms}{km\,s$^{-1}$}
\newcommand{\ebv}{E(B$-$V)}
\newcommand{\msun}{M$_\odot$}
\newcommand{\sulyn}{SU~Lyn}

\title[SU Lyncis: a hard X-ray bright symbiotic star]
      {SU Lyncis, a hard X-ray bright M giant: Clues point to a large hidden population of symbiotic stars}

\author[K. Mukai et al.]{K. Mukai,$^{1,2}$\thanks{E-mail: Koji.Mukai@nasa.gov (KM)}
G.J.M. Luna,$^{3}$
G. Cusumano,$^{4}$
A. Segreto,$^{4}$
U. Munari,$^{5}$
J.L. Sokoloski,$^{6}$
\newauthor
A.B. Lucy,$^{6}$
T. Nelson$^{7}$
and N.E. Nu\~nez$^{8}$
\\
$^{1}$CRESST and X-ray Astrophysics Laboratory, NASA Goddard Space Flight Center, Greenbelt, MD 20771, USA\\
$^{2}$Department of Physics, University of Maryland, Baltimore County, 1000 Hilltop Circle, Baltimore, MD 21250, USA\\
$^{3}$Instituto de Astronom\'ia y F\'isica del Espacio (IAFE, CONICET-UBA), Av. Inte. G\"uiraldes 2620, C1428ZAA, Buenos Aires, Argentina\\
$^{4}$INAF -- Istituto di Astrofisica Spaziale e Fisica Cosmica, Via U. La Malfa 153, I-90146 Palermo, Italy\\
$^{5}$INAF -- Astronomical Observatory of Padova, 36012 Asiago (VI), Italy\\
$^{6}$Columbia Astrophysics Laboratory, Columbia University, New York, NY 10027, USA\\
$^{7}$Department of Physics and Astronomy, University of Pittsburgh, 3941 O'Hara St, Pittsburgh, PA 15260, USA\\
$^{8}$Instituto de Ciencias Astron\'omicas de la Tierra y del Espacio (ICATE-UNSJ, CONICET), Av. Espa\~na (S) 1512, J5402 DSP, San Juan, Argentina\\
}

\date{Accepted XXX. Received YYY; in original form ZZZ}

\pubyear{2016}

\begin{document}
\label{firstpage}
\pagerange{\pageref{firstpage}--\pageref{lastpage}}
\maketitle

\begin{abstract}
Symbiotic star surveys have traditionally relied almost exclusively
on low resolution optical spectroscopy. However, we can obtain a more
reliable estimate of their total Galactic population by using all available
signatures of the symbiotic phenomenon. Here we report the discovery of a
hard X-ray source, 4PBC~J0642.9+5528, in the \swift\ hard X-ray all-sky
survey, and identify it with a poorly studied red giant, \sulyn, using pointed
\swift\ observations and ground-based optical spectroscopy. The X-ray
spectrum, the optical to UV spectrum, and the rapid UV variability of
\sulyn\ are all consistent with our interpretation that it is a symbiotic
star containing an accreting white dwarf. The symbiotic nature
of \sulyn\ went unnoticed until now, because it does not exhibit emission
lines strong enough to be obvious in low resolution spectra. We argue that
symbiotic stars without shell-burning have weak emission lines, and that
the current lists of symbiotic stars are biased in favor of shell-burning
systems. We conclude that the true population of symbiotic stars has been
underestimated, potentially by a large factor.
\end{abstract}




\section{Introduction}

The traditional, phenomenological definition of a symbiotic star is a late
type giant with a companion, hot enough to result in prominent high-excitation
emission lines in the optical \citep{Kenyon1986}. While we know of several
symbiotic X-ray binaries with neutron star accretors (see, e.g.,
\citealt{Masetti2007,Corbet2008,SymbXB}), and symbiotic stars with main sequence
accretors may exist, the majority of symbiotic stars appear to be wide
binaries in which white dwarfs accrete from red giants.  

In recent years, we have gained new insights into the symbiotic phenomenon
through X-ray observations.  First was the discovery of four symbiotic stars,
CH~Cyg, T~CrB, RT~Cru, and V648~Car (also known as SS73 17 and CD $-$57 3057),
in the \swift\ Burst Alert Telescope (BAT) survey of the 14--195 keV sky
(\citealt{Kennea2009}; see also \citealt{LS2007,Luna2008,Smith2008}). The
hard X-ray spectra of these sources can be described as optically thin, thermal
emission with a high temperature, and are interpreted as due to accretion
onto the white dwarf surface via the boundary layer. Subsequently, pointed
\swift\ X-Ray Telescope (XRT) survey of known symbiotic stars
(\citealt{Luna2013} and references therein) showed that many other symbiotic
stars had a similar X-ray emission component, although not luminous and/or
hard enough to be detectable in the BAT survey. \cite{Luna2013} called this
type of X-ray emission $\delta$-type, expanding the earlier classification
scheme of \cite{Muerset1997}. The $\delta$-type X-ray component is often
highly absorbed and cannot be detected in the soft X-rays. The column density
of the absorber often far exceeds that expect from the interstellar medium
(ISM), and is seen to be variable from observation to observation: the
absorber is intrinsic to the system.

These observations allow us to divide symbiotic stars into shell-burning
systems, powered by nuclear fusion as well as accretion, and non-burning
systems, powered by accretion alone. There is no known instance of a
shell-burning system in which a $\delta$-type X-ray emission is detected.
In addition, the \swift\ UV and Optical Telescope
(UVOT) data show a high degree of UV variability in the non-burning
symbiotic stars, interpreted as flickering of the accretion disc,
while shell-burning symbiotics have steady UV emission \citep{Luna2013}.

\cite{Kennea2009} showed that 4 out of 461 BAT sources in the 22-month
survey catalogue \citep{BAT22m} were non-burning symbiotic stars.
In this Letter, we report the first discovery since then of a new
symbiotic star among BAT sources.

\section{Observations and Data Analysis}

\subsection{BAT Survey Detection and Analysis}

\begin{figure}
	\includegraphics[width=\columnwidth]{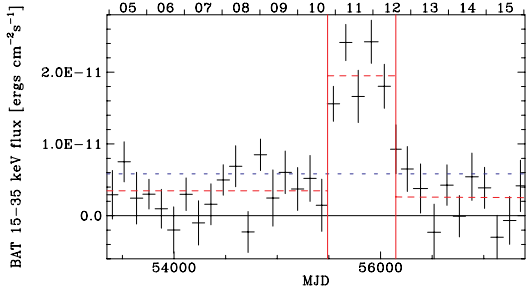}
    \caption{The 11-year 15--35 keV BAT light curve of \sulyn\ in
    		 120 day bins. Vertical red lines indicate the high state
		 (2010 October 14 to 2012 August 1) as identified by
		 the survey pipeline. The short dashed blue line is the
		 overall weighted average, while long dashed red lines show
		 the weighted average during the high state and the periods
		 before and after.}
    \label{fig:BATlc}
\end{figure}

The \swift\ mission \citep{SwiftRef} performs a sensitive all-sky survey
of the hard X-ray (14--195 keV) sky as a by-product of its main objective
to detect and observe Gamma-ray bursts.  GC and AS have been engaged in
an effort to produce a series of catalogues of BAT sources (see, e.g.,
\citealt{Palermo54}) using the BAT\_Imager code \citep{Segreto2010}, and
are preparing the fourth version of the catalogue covering the first 100
months of the mission (2004 December to 2013 April) and containing 1710
sources\footnote{http://bat.ifc.inaf.it/100m\_bat\_catalog/100m\_bat\_catalog\_v0.0.htm}.
One new source in this catalogue is 4PBC~J0642.9+5528.

The batch processing for the catalogue indicated that this source had
its highest signal-to-noise (S/N) in the 15--35 keV range, and during
2010 October 14 to 2012 August 1 (hereafter `high state').  The survey
products include the average and the high state spectra of
4PBC~J0642.9+5528. To investigate its variability further, we extracted
15--35 keV light curve over the period 2004 December 8 to 2016 January 11
in 15 day bins, in which we converted BAT count rate to flux using
a factor of 3.147$\times 10^{-7}$ \epcs\ per 1 \cps, appropriate for
the spectral shape of \sulyn; we show the light curve rebinned into
120 day bins in Figure\,\ref{fig:BATlc}.
 
The only previously known X-ray source in the BAT error circle is
1RXS~J064255.9+552835, and the only known optical object within
its error circle is \sulyn, listed in the General Catalog of Variable
Stars as a semi-regular variable (magnitude range: 10.5--9.6).  There do
not appear to be any publications that discuss in-depth studies of \sulyn.
A single M giant is a very weak, almost undetectable, {\sl soft\/} X-ray
source at most (see, e.g., \citealt{Ayres2003}), and never a hard ($>$10
keV) X-ray source. We therefore arranged for follow-up \swift\ and optical
observations to confirm this association.

\subsection{Pointed \swift\ Observations}

\begin{table}
	\centering
	\caption{Pointed \swift\ Observations of \sulyn}
	\label{tab:SwiftObs}
	\begin{tabular}{lcrl} 
		\hline
		ObsID & Date & XRT Exposure (s) & UVOT Filter \\
		\hline
                00034150001 & 2015-11-20 & 2963 & UVM2 \\
                00085853001 & 2015-11-30 & 2274 & U    \\
                00085853002 & 2015-12-05 &  649 & UVW2 \\
                00085853003 & 2015-12-07 &  318 & UVW1 \\
		\hline
	\end{tabular}
\end{table}

KM and GC each requested a 3 ks \swift\ observation to confirm the
association of 4PBC~J0642.9+5528 with \sulyn, and these observations
were carried out between 2015 November 20 and December 7 (see
Table\,\ref{tab:SwiftObs}). During observation 0003415001, UVOT data
were taken with the UVM2 filter in event mode; the other observations
took UVOT image mode data in U, UVW2 and UVW1 filters.

The \swift\ data were analyzed using HEAsoft version 6.18 with the
latest calibration files. We first combined the XRT event files from
the four observations.
The spectrum and light
curve of the sources were extracted from a circular region with a radius
of 20 pixels ($\sim 47''$) centred on the SIMBAD coordinates of
\sulyn, and the background from an annular region with the inner and
the outer radii of 27 and 54 pixels, respectively. We used the response
file \texttt{swxpc0to12s6\_20130101v014.rmf} in the calibration
database, and used \texttt{xrtmkarf} to construct the ancillary response
file. In our timing analysis, we used the \texttt{xrtlccorr} tool
to correct for the presence of dead columns.

We used the
\texttt{uvotsource} tool to extract the magnitudes from
the UVOT image mode data, with a source aperture of 5$''$ radius.
We discarded the U band image, since \sulyn\ was found to be saturated.
In the UV filters, \sulyn\ suffered some coincidence losses that were
correctable. We used the same aperture to extract the light curves from
the event mode data using \texttt{uvotevtlc}, which applied the corrections
for coincidence losses and large-scale sensitivity variations \citep{UVOTCal}.

\subsection{Optical Spectroscopy}

One of us (UM) initiated a campaign to obtain optical spectra of \sulyn, and
selected field stars around it, using two telescopes at Asiago.
We use the 1.82m telescope with the Echelle spectrograph to obtain high
resolution spectra covering 3650--7350\AA\ with a resolving power of
20,000 in 30 orders without inter-order gaps. We use the 1.22m telescope
with the Boller and Chivens spectrograph  to obtain low resolution spectra
covering 3300--7850\AA\ with 2.31\AA\ per pixel and a full width half maximum
of 2.2 pixels.  From this on-going campaign, we describe in this letter
a subset of results obtained between 2015 October and 2016 January to
provide the essential context for interpreting the X-ray and UV observations.

\section{Results}

In the \swift\ XRT data, there is one and only one 2--10 keV X-ray
source within the error circle of 4PBC~J0642.9+5528,
1.3$''$ from the SIMBAD position of \sulyn\ with an error
circle of 3.8$''$ radius. There is one UVOT source within the XRT
error circle, 0.2$''$ from \sulyn\ with an error circle of 0.42$''$ radius:
\sulyn\ almost certainly is the counterpart of the BAT source.

\subsection{X-ray Variability}

Figure\,\ref{fig:BATlc} clearly demonstrates the long-term variability
of \sulyn. The average BAT 15--35 keV band flux was (0.58$\pm$0.05)
$\times 10^{-11}$ \epcs. The high state flux was (1.95$\pm$0.12)
$\times 10^{-11}$ \epcs, while the normal states flux was at
(0.35$\pm$0.07) and (0.26$\pm$0.10) $\times 10^{-11}$ \epcs,
respectively, before and after the high state. The high state flux was
$\sim$6.5 times higher than in normal state, and elevated the overall average
by a factor of $\sim$2.

We have calculated the average \swift\ XRT count rates for the
four observations in the hard (2--10 keV) and the soft (0.3--2 keV)
bands. While the hard rate stayed approximately constant at 0.04$\pm$0.01
\cps, the soft rate varied from (2.2$\pm$1.1)$\times 10^{-3}$ \cps\ in the
first observation to 0.05$\pm$0.01 \cps\ in the last, implying variable
X-ray absorption.

\subsection{X-ray Spectrum}

\begin{figure}
	\includegraphics[width=\columnwidth]{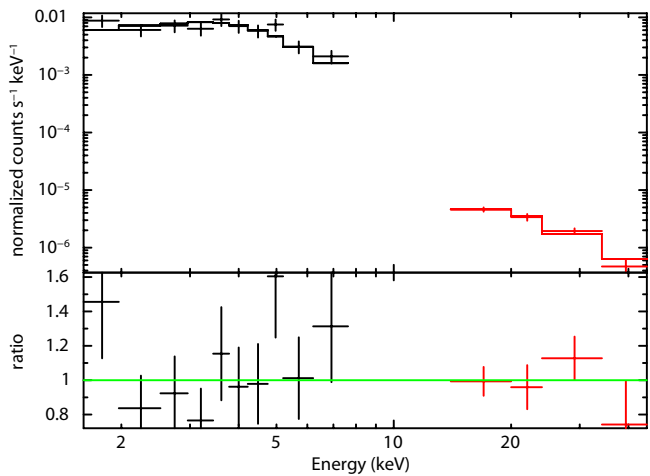}
        \caption{The joint fit of the BAT high-state data and 2016
          November/December XRT data for \sulyn. See text for details.}
    \label{fig:xrspec}
\end{figure}

We combined all XRT data in our spectral analysis because spectra from
individual observations were noisy.  We fit the average XRT spectrum
simultaneously with the BAT high-state spectrum, allowing for a
cross-normalization factor to account for the long-term variability,
which we assume to be energy independent.  In one fit, we used an
\texttt{APEC} thermal plasma model with \texttt{tbabs} absorber model
and obtained \nh\ =2.9$^{+1.1}_{-0.9} \times 10^{22}$ cm$^{-2}$ and
kT=17$^{+6}_{-4}$ keV. The BAT cross-normalization constant was 7.7,
and the inferred 0.3--50 keV unabsorbed flux during the XRT observations
was 1.05$\times 10^{-11}$ \epcs. Given the spectral shape, we expect
little additional flux outside this band. We show this fit in
Figure\,\ref{fig:xrspec}. We have also fitted the data using the cooling
flow model, \texttt{mkcflow}. A similar quality fit was obtained, with
the maximum temperature ($kT_{max}$) of $26^{+12}_{-7}$ keV. When we used
the average BAT spectrum instead, again with the average XRT spectrum,
the results were similar except that the cross-normalization factor was
$\sim$2, and the parameter values had larger errors. While better data
are necessary for a definitive fit, these results are consistent with those
on well-studied $\delta$-type symbiotic stars (see, e.g.,
\citealt{LS2007,Smith2008}). 

\subsection{UV Variability}

\begin{figure}
	\includegraphics[width=\columnwidth]{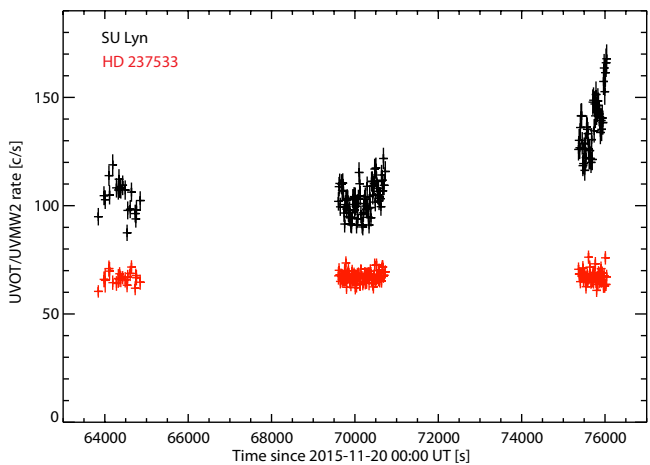}
        \caption{The UVOT UVM2 band light curve of \sulyn\ in 15 s bins.
		 We also show that of HD~237533 for comparison.}
    \label{fig:uvotlc}
\end{figure}

We used the UVM2 event mode data to investigate the UV variability of \sulyn.
We extracted the light curves of both \sulyn\ and
another bright star in the UVOT field-of-view, HD 237533, in 15 second bins,
and show them in Figure\,\ref{fig:uvotlc}. \sulyn\ is clearly highly variable
both within individual \swift\ orbits, and from one \swift\ orbit to the next.
To quantify the degree of variability, we calculated the fractional
RMS variability ($s_{frac}$) and also compared the measured light curve
standard deviation ($s$) with the average error bar ($s_{exp}$). We find
$s_{frac}$ of 7\% and $s/s_{exp}$=1.73 during the first orbit;
7\% and 1.69 during the second; and 10\% and 2.42 during the third.

We also analyzed archival {\sl GALEX\/} data. \sulyn\ had NUV
(1771--2831\AA) and FUV (1350--1780\AA) magnitudes of 17.42$\pm$0.02
and 16.64$\pm$0.03 on 2006 December 21, and 15.80$\pm$0.02 and 16.11$\pm$0.03
on 2007 January 27. That is, \sulyn\ was much fainter and variable
during the {\sl GALEX} observations.

\subsection{Optical to UV Spectrum of \sulyn}

\begin{figure}
	\includegraphics[angle=270,width=\columnwidth]{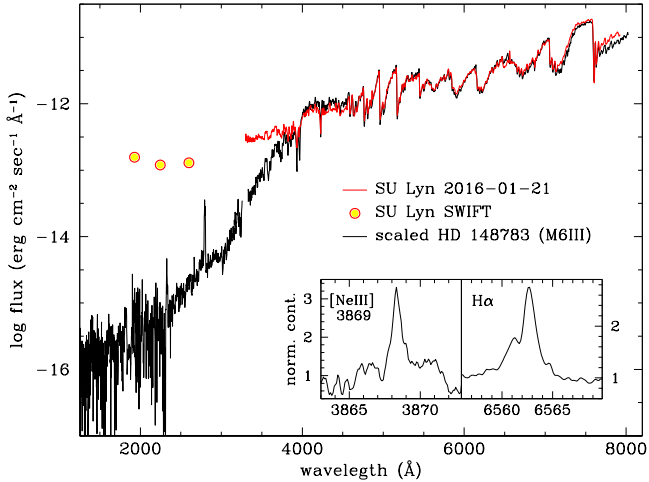}
        \caption{Low resolution optical spectrum of \sulyn\ with \swift\ UVOT
          points, compared with optical and IUE spectra of MIII star, HD~14783.
	  Insets show the H$\alpha$ and [NeIII] 3869 profiles from the
	  high resolution spectra taken on the same night.}
    \label{fig:spectrum}
\end{figure}

A comparison of our low resolution optical spectrum of \sulyn, obtained
on 2016 January 21, with that of HD 148783, the MKK spectral standard
for M6III, highlights the evidence for mass transfer.  As shown in
Figure\,\ref{fig:spectrum}, they match nicely longward of 4000\AA,
suggesting that this is close to the spectral type of the giant in \sulyn.
We investigated this further by fitting a series of templates
constructed by linearly interpolating M5III and M6III spectra taken from
the atlas of \cite{MgiantAtlas}. We found a near-perfect match between
\sulyn\ and an M5.8III cool giant subjected to a reddening of \ebv\ =0.07
following a standard R$_V$=3.1 extinction law. The allowed range of the
spectral type is found to be between M5.6 and M5.9, correlated with
the $\pm$0.02 uncertainty on \ebv.

\sulyn\ has a Tycho-2 V magnitude of 8.642, which equals V=8.46 in
the standard Johnson system \citep{Bessell2000}. For the M5.8III
spectral type, the R$_V$=3.1 reddening law implies
A$_V$=3.75\ebv\ =0.263 \citep{FM2003}. The
absolute magnitude of an M5.8III star is M$_V = -0.83$ on the
\cite{Sowell2007} scale. Thus we estimate the distance to \sulyn\ to
be 640 pc. The dominant source of uncertainty is likely to be the accuracy of
the \cite{Sowell2007} scale, to which we assign $\pm$0.25 mag. The
resulting uncertainty on the distance is $\sim$10\%.

Our Echelle spectra of \sulyn\ show that the each line of the NaI D
doublet consists of 3 components.  For comparison, we have collected
the spectra of five hot stars (which therefore do not show intrinsic
NaI lines) within 1$^\circ$ of the position on the sky of \sulyn.  We also
observed a sixth, cooler star (HD~237529), with a radial velocity
high enough to decouple the stellar NaI doublet from the interstellar one.
This allowed us to identify the component with the lowest heliocentric
velocity ($-$4.1 \kms) in the \sulyn\ spectra as interstellar.

\begin{figure}
	\includegraphics[width=\columnwidth]{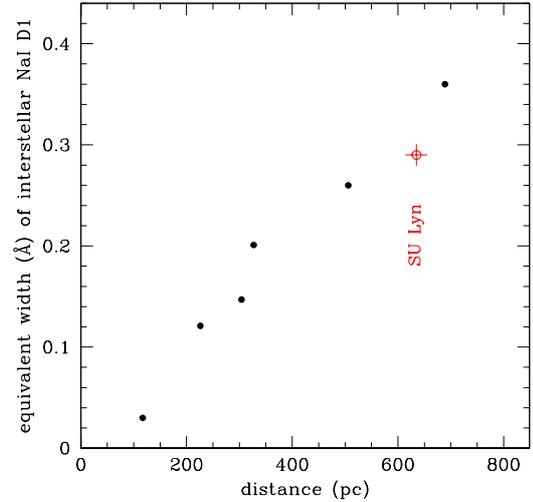}
        \caption{The equivalent widths of the interstellar NaI line at
          5889.953\AA\ of \sulyn\ and six field stars, plotted
          against their estimated distances.}
    \label{fig:nai}
\end{figure}

We present the equivalent width of interstellar NaI D1 line (at
5889.953\AA) in \sulyn\ and these six field stars in Table\,\ref{tab:NaD}
and in Figure\,\ref{fig:nai}.
The stars are listed from the nearest to most distant, estimated using
Hipparcos and spectroscopic parallaxes.  On the one hand, the equivalent
width (EqW) we measure for \sulyn\ of 0.290\AA\ would suggest
\ebv\ =0.109, somewhat larger than the value we obtained from
the template fit, using the \cite{MZ1997} calibration for the
ISM.  On the other hand, it nicely fits with the
progression of the field stars from HD 46608 (117 pc, EqW=0.030\AA)
to HD 237549 (689 pc, EqW=0.360 \AA), suggesting that our distance
estimate is reasonably accurate. Considering both the template fitting
and NaI D1 line results, we conservatively estimate the distance
to \sulyn\ to be 640$\pm$100 pc.

Finally, we find clear evidence for UV emission powered by binary
interaction by combining low resolution optical spectrum and our \swift\ UVOT
data converted to fluxes using the calibration of \cite{UVOTCal}.
While we find an excellent match between the optical spectrum of \sulyn\ and
the scaled optical spectrum of HD~148783, the UVOT data of \sulyn\ are more
than 2 orders of magnitude above the scaled archival IUE spectra of HD~14783
(Figure\,\ref{fig:spectrum}). Moreover, we observe variable emission lines
of hydrogen Balmer series, [NeIII], and CaII in our high resolution data.
That is, \sulyn\ is a symbiotic star.

\section{Discussion}

\begin{table}
	\centering
	\caption{The distance (from Hipparcos and spectroscopic parallaxes)
	         and interstellar Na D1 line equivalent widths of \sulyn\ and
		 6 field stars}
	\label{tab:NaD}
	\begin{tabular}{llrl} 
		\hline
		Star        & Spectral Type & Distance (pc) & Na D1 Eq.W (\AA) \\
		\hline
                HD 46608    & A2.5V   & 117 & 0.030 \\
                HD 48841    & A1V     & 226 & 0.121 \\
                HD 237526   & A2V     & 304 & 0.147 \\
                HD 46732    & A0V     & 327 & 0.201 \\
                HD 237529   & K2III   & 506 & 0.260 \\
                \sulyn\     & M5.8III & 640 & 0.290 \\
                HD 237549   & B8V     & 689 & 0.360 \\
		\hline
	\end{tabular}
\end{table}

We have shown that \sulyn, previously catalogued as a semi-regular
variable, is in fact a hard X-ray luminous symbiotic star. We use
our \swift\ results to infer the nature of the hot component as
follows. Assuming a distance of 640 pc, the X-ray luminosity
was $\sim$5.2 $\times 10^{32}$ \eps\ during the XRT observations,
and $\sim$4.0 $\times 10^{33}$ \eps\
during the high state. We estimate the UV flux to be about 1.2$\times 10^{-13}$
ergs\,cm$^{-2}$s$^{-1}$\AA$^{-1}$, for a total UV (2000--4000\AA)
flux of 2.4$\times 10^{-10}$ \epcs, or UV luminosity of 1.2$\times 10^{34}$
\eps. Thus, the UV luminosity exceeds the X-ray luminosity by a
substantial margin, even before applying a bolometric correction.
This strongly argues against a symbiotic X-ray binary interpretation,
in which the accreting object is a neutron star, since accreting neutron
stars release most of the gravitational potential energy in the X-rays.

The X-ray and UV luminosities of \sulyn\ indicate that it contains
an accreting white dwarf. Moreover, when compared to the nine
known symbiotic stars whose $\delta$-type X-ray emission was discovered by
\cite{Luna2013}, \sulyn\ appears more X-ray luminous than at least five.
In the UVOT event mode data on \sulyn, we have found a strong UV variability
down to a sub-minute time scale
for the first time in such a system. The observed variability
was of the same order as that obtained by \cite{Luna2013} in several
other symbiotic stars, although they were only sensitive to variability
on longer time scales since they used UVOT image mode data.
\cite{Luna2013} found that such strong UV variability is a characteristic
of non-burning symbiotic stars. Therefore, we infer that \sulyn\ is also
powered purely by accretion, with the observed UV luminosity predominantly
from the accretion disc.

We can constrain the white dwarf mass in \sulyn\ using our X-ray spectral
fits, because the white dwarf mass determines the maximum shock temperature
that can be reached in an optically thin boundary layer \citep{Byckling2010}.
The observed $kT_{max}$ is consistent with a 1 \msun\ white dwarf (with a
predicted $kT_{max}$ of 28 keV). In this scenario, the observed X-ray
luminosity implies an accretion rate of 7.0 $\times 10^{-11}$ \msun\,yr$^{-1}$.
The observed UV luminosity corresponds to an accretion rate of
1.6$\times 10^{-9}$ \msun\,yr$^{-1}$, assuming that it is entirely from
the accretion disc, and is likely to be higher due to bolometric correction.
Thus, although the boundary layer could be optically thin if an additional
source of UV luminosity such as shocks is present, the rapid UV variability
constrains the possible nature of, or the contribution by, such a source.

Alternatively, the boundary layer in \sulyn\ may be partially optically
thick, which is thought to require an accretion rate above
$\sim$1.0$\times 10^{-8}$ \msun\,yr$^{-1}$ (see \citealt{Natalia2016}
and references therein), and hence a factor of $>$6 bolometric correction
for the disk UV luminosity. Such a boundary layer emit a residual optically
thin X-ray emission with a lower luminosity and with a lower temperature
than the pure optically thin case (see, e.g., \citealt{WMM2003}).
In this interpretation, the white dwarf in \sulyn\ is likely to be more
massive than $\sim$1 \msun.


The manner of discovery of \sulyn\ is very similar to that of 4~Dra,
whose hot component was discovered in the UV \citep{Reimers1985}.
Later observations established this as a symbiotic star with a
$\delta$-type X-ray emission (\citealt{Wheatley2003,Natalia2016}).
The symbiotic nature of \sulyn\ and 4~Dra were not recognized from the ground
because neither has emission lines strong enough to show up in low
resolution surveys. In typical symbiotic stars, prominent emission lines
arise when the red giant wind is photoionized by a central hot source
which is thought to be a $\sim$10$^5$K blackbody-like source with
luminosities 100--10000 L$_\odot$ (see, e.g., \citealt{Skopal2015}),
or $4 \times 10^{35}$--$4 \times 10^{37}$ \eps.  The UV luminosity
of T~CrB is estimated to be lower, and the optical
emission lines are prominent only some of the time (see, e.g.,
\citealt{Munari2016}). The UV luminosity of \sulyn\ may well be lower
still, similar to that of 4~Dra (2.6$\times 10^{34}$ \eps;
\citealt{Skopal2015}). It is therefore reasonable that the
resulting emission lines in the optical are too weak to show up in
low resolution spectroscopic surveys.

The large difference in the hot component luminosity, in turn, is due
to the different energy sources: nuclear fusion vs. accretion. The
former produces $\sim$50 times more energy per nucleon than the latter.
The shell-burning symbiotic stars exhibit prominent emission lines;
non-burning ones have weak lines at most.  The existing catalogues
of symbiotic stars rely heavily on the emission lines and are therefore
biased in favor of shell-burning symbiotics.


\sulyn\ and 4~Dra are two examples of non-burning symbiotics
without prominent lines, and other similar systems likely exist.
This hidden population is potentially large. The BAT detection of
\sulyn\ was possible because of its relative proximity, and because
of its X-ray high state. Without the high state, the BAT survey
can only detect \sulyn\ out to $<$500 pc. The \rosat\ all-sky survey
could only detect 4~Dra out to $<$200 pc. Scaling by the area of the
Galactic disc, these suggest of order 4 \sulyn-like systems and of order
25 4~Dra-like systems within 1 kpc, implying a higher space density than
that of known symbiotic stars. The three hard X-ray emitting, UV excess
AGB stars discovered by \cite{Sahai2015} are likely to be part of this
same population.

\cite{MR1992} estimated the total Galactic population of symbiotic stars to
be up to $\sim$3 $\times 10^5$, up from the previous estimate of
$\sim$3000. Both estimates are extrapolation from the $\sim$150
systems known at the time, but with different estimates of their distances.
We believe it is quite possible that we may face another significant
revision in the total Galactic population of symbiotic stars, once
a survey for \sulyn-like systems can be performed.

\section{Conclusions}

We have discovered that \sulyn, previously catalogued as a semi-regular
variable, is a hard X-ray source in the \swift\ BAT survey catalogue. Based
on the \swift\ and ground-based data, we interpret \sulyn\ as a symbiotic
star powered purely by accretion onto a white dwarf. The lack of shell
burning leads to \sulyn\ having very weak symbiotic signatures in the optical.
Since existing catalogues of symbiotic stars rely heavily on prominent emission
lines, non-burning symbiotics are likely severely under-represented
in these catalogues.  Further observations of \sulyn\ are highly desirable
for us to tune our search strategy for other members of this hidden
population of symbiotic stars.

\section*{Acknowledgements}

We thank Neil Gehrels, the PI of the \swift\ mission, for a generous
allocation of TOO time. GJML and NEN are members of the `Carrera del
Investigador Cientifico (CIC)' of CONICET and acknowledge support from
Argentina under grant ANPCYT-PICT 0478/14. JLS and ABL acknowledge support
from NASA ADAP grant NNX15AF19G. JLS thanks Scott Kenyon for conversations
(a decade ago) about shell burning and selection bias.





\begin{thebibliography}{99}

\bibitem[\protect\citeauthoryear{Ayres, Brown \& Harper}{2003}]{Ayres2003}
Ayres T.R., Brown A., Harper, G.M., 2003, ApJ, 598, 610

\bibitem[\protect\citeauthoryear{Bessell}{2000}]{Bessell2000}
Bessell M. S., 2000, PASP, 112, 961

\bibitem[\protect\citeauthoryear{Byckling et al.}{2010}]{Byckling2010}
Byckling K., Mukai K., Thorstensen J.R., Osborne J.P., MNRAS,
408, 2298


\bibitem[\protect\citeauthoryear{Corbet et al.}{2008}]{Corbet2008}
Corbet R.H.D., Sokoloski J.L., Mukai K., Markwardt C.B., Tueller, J.,
2008, ApJ, 675, 1424

\bibitem[\protect\citeauthoryear{Cusumano et al.}{2010}]{Palermo54}
Cusumano G. et al., 2010, A\&A, 524, A64


\bibitem[\protect\citeauthoryear{Enoto et al.}{2014}]{SymbXB}
Enoto T. et al., 2014, ApJ, 786, A127

\bibitem[\protect\citeauthoryear{Fiorucci \& Munari}{2003}]{FM2003}
Fiorucci M. Munari U. 2003, A\&A, 401, 781

\bibitem[\protect\citeauthoryear{Fluks et al.}{1994}]{MgiantAtlas}
Fluks M.A., Plez B., The P.S., de Winter D., Westerlund B.E.,
Steenman H.C., 1994, A\&AS, 105, 311

\bibitem[\protect\citeauthoryear{Gehrels et al.}{2004}]{SwiftRef}
Gehrels N. et al., 2004, ApJ, 611, 1005

\bibitem[\protect\citeauthoryear{Kennea et al.}{2009}]{Kennea2009}
Kennea J.A., Mukai K., Sokoloski J.L., Luna G.J.M., Tueller J.,
Markwardt C.B., Burrows D.N, 2009, ApJ, 701, 1992

\bibitem[\protect\citeauthoryear{Kenyon}{1986}]{Kenyon1986}
Kenyon S.J., 1986, The symbiotic stars. Cambridge Univ. Press,
Cambridge, UK

\bibitem[\protect\citeauthoryear{Luna \& Sokoloski}{2007}]{LS2007}
Luna G.J.M., Sokoloski J.L., 2007, ApJ, 671, 741

\bibitem[\protect\citeauthoryear{Luna, Sokoloski \& Mukai}{2008}]{Luna2008}
Luna G.J.M., Sokoloski J.L., Mukai K., 2008, ASPC, 401, 342

\bibitem[\protect\citeauthoryear{Luna et al.}{2013}]{Luna2013}
Luna G.J.M., Sokoloski J.L., Mukai K., Nelson T., 2013,
A\&A, 559, A6

\bibitem[\protect\citeauthoryear{Masetti et al.}{2007}]{Masetti2007}
Masetti N. et al., 2007, A\&A, 470, 331

\bibitem[\protect\citeauthoryear{Muerset, Wolff \& Jordan}{1997}]{Muerset1997}
Muerset U., Wolff B., Jordan S., 1997, A\&A, 319, 201

\bibitem[\protect\citeauthoryear{Munari \& Renzini}{1992}]{MR1992}
Munari U., Renzini A., 1992, ApJ, 397, L87

\bibitem[\protect\citeauthoryear{Munari \& Zwitter}{1997}]{MZ1997}
Munari U., Zwitter T., 1997, A\&A, 318, 269

\bibitem[\protect\citeauthoryear{Munari, Dallaporta \& Cherini}{2016}]{Munari2016}
Munari U., Dallaporta S., Cherini, G., 2016, NewA, 47, 7



\bibitem[\protect\citeauthoryear{Nu\~nez et al.}{2016}]{Natalia2016}
Nu\~nez N.E., Nelson T., Mukai K., Sokoloski J.L., Luna G.J.M.,
2016, ApJ, in press (arXiv:1604.05980)

\bibitem[\protect\citeauthoryear{Poole et al.}{2008}]{UVOTCal}
Poole T.S. et al., 2008, MNRAS, 383, 627

\bibitem[\protect\citeauthoryear{Reimers}{1985}]{Reimers1985}
Reimers D. 1985, A\&A, 142, L16

\bibitem[\protect\citeauthoryear{Sahai et al.}{2015}]{Sahai2015}
Sahai, R., Sanz-Forcada, J., S\'anchez Contreras, C., Stute, M.,
2015, ApJ, 810, A77



\bibitem[\protect\citeauthoryear{Segreto et al.}{2010}]{Segreto2010}
Segreto A., Cusumano G., Ferrigno C., La Parola V., Mangano V.,
Mineo T., Romano P., 2010, A\&A, 510, A47


\bibitem[\protect\citeauthoryear{Smith et al.}{2008}]{Smith2008}
Smith R.K., Mushotzky R., Mukai K., Kallman T., Markwardt C.B.,
Tueller J. 2008, PASJ, 60, S43

\bibitem[\protect\citeauthoryear{Sowell et al.}{2007}]{Sowell2007}
Sowell J. R., Trippe M., Caballero-Nieves S. M., Houk N., 2007, AJ 134, 1089

\bibitem[\protect\citeauthoryear{Skopal}{2015}]{Skopal2015}
Skopal A., 2015, NewA, 36, 116

\bibitem[\protect\citeauthoryear{Tueller et al.}{2010}]{BAT22m}
Tueller J. et al., 2010, ApJS, 186, 378

\bibitem[\protect\citeauthoryear{Wheatley, Mauche \& Mattei}{2003}]{WMM2003}
Wheatley P.J., Mauche C.W., Mattei J.A., 2003, MNRAS, 345, 49

\bibitem[\protect\citeauthoryear{Wheatley, Mukai \& de Martino}{2003}]{Wheatley2003}
Wheatley P.J., Mukai K., de Martino D., 2003, MNRAS, 346, 855

\end{thebibliography}




%
%


\bsp	
\label{lastpage}
\end{document}